\documentclass[12pt,preprint]{aastex}

\usepackage{epsfig}
\usepackage{amsmath}
\usepackage{amssymb}
\usepackage{amsbsy}
\usepackage{placeins}
\usepackage{color}
\usepackage{lscape}

\newcommand{\D}{\mbox{d}}

\newcommand{\iles}{{\sc iles}}

\newcommand{\pdf}{{PDF}}
\newcommand{\jpdf}{{JPDF}}

\newcommand{\Ka}{\mathrm{Ka}}
\newcommand{\Da}{\mathrm{Da}}

\usepackage{ulem}

\newcommand{\gtaprx}{\lower .1ex\hbox{\rlap{\raise .6ex\hbox{\hskip .3ex
        {\ifmmode{\scriptscriptstyle >}\else
                {$\scriptscriptstyle >$}\fi}}}
        \kern -.4ex{\ifmmode{\scriptscriptstyle \sim}\else
                {$\scriptscriptstyle\sim$}\fi}}}
\newcommand{\ltaprx}{\lower .1ex\hbox{\rlap{\raise .6ex\hbox{\hskip .3ex
        {\ifmmode{\scriptscriptstyle <}\else
                {$\scriptscriptstyle <$}\fi}}}
        \kern -.4ex{\ifmmode{\scriptscriptstyle \sim}\else
                {$\scriptscriptstyle\sim$}\fi}}}

\usepackage{epsf,color,dcolumn}

\newcolumntype{d}{D{.}{.}{-1}}

\epsfverbosetrue

\begin{document}

\title{Distributed Flames in Type Ia Supernovae}

\author{A.~J.~Aspden\altaffilmark{1}, J.~B.~Bell\altaffilmark{1}, and S.~E.~Woosley\altaffilmark{2}}

\altaffiltext{1}{Lawrence Berkeley National Laboratory, 1 Cyclotron
Road, MS 50A-1148, Berkeley, CA 94720}
\altaffiltext{2}{Department of Astronomy and Astrophysics, University
of California at Santa Cruz, Santa Cruz, CA 95064}

\begin{abstract}
At a density near a few $\times 10^7$ g cm$^{-3}$, the subsonic burning
in a Type Ia supernova enters the distributed regime (high Karlovitz
number).  In this regime, turbulence disrupts the internal structure
of the flame, and so the idea of laminar burning propagated by
conduction is no longer valid.  The nature of the burning in this
distributed regime depends on the turbulent Damk\"ohler number
($\Da_T$), which steadily declines from much greater than one to less
that one as the density decreases to a few $\times 10^6$ g cm$^{-3}$.
Classical scaling arguments predict that the turbulent flame speed
$s_T$, normalized by the turbulent intensity $\check{u}$, follows
$s_T/\check{u}=\Da_T^{1/2}$ for $\Da_T\ltaprx 1 $.  The flame in this
regime is a single turbulently-broadened structure that moves at a
steady speed, and has a width larger than the integral scale of the
turbulence.  The scaling is predicted to break down at
$\Da_T\approx1$, and the flame burns as a turbulently-broadened 
effective unity Lewis number flame.  This flame burns locally with
speed $s_\lambda$ and width $l_\lambda$, and we refer to this kind of
flame as a $\lambda$-flame.  The burning becomes a collection of
$\lambda$-flames spread over a region approximately the size of the
integral scale.  While the total burning rate continues to have a
well-defined average, $s_T \sim \check{u}$, the burning is
unsteady.  We present a theoretical framework, supported by 
both 1D and 3D numerical simulations, for the burning in these two
regimes.  Our results indicate that the average value of $s_T$ can
actually be roughly twice $\check{u}$ for $\Da_T\gtaprx 1$, and that
localized excursions to as much as five times $\check{u}$ can occur.
We also explore the properties of the individual flames,
which could be sites for a transition to detonation when $\Da_T\sim1$.
The $\lambda$-flame speed and width can be predicted based on the
turbulence in the star (specifically the energy dissipation rate
$\varepsilon^*$) and the turbulent nuclear burning time scale of the
fuel $\tau_{\mathrm{nuc}}^T$.  We propose a practical method for
measuring $s_\lambda$ and $l_\lambda$ based on the scaling relations
and small-scale computationally-inexpensive simulations.  This
suggests that a simple turbulent flame model can be easily constructed
suitable for large-scale distributed supernovae flames.  These results
will be useful both for characterizing the deflagration speed in
larger full-star simulations, where the flame cannot be resolved, and
for predicting when detonation occurs.
\end{abstract}


\section{INTRODUCTION}
\label{sec:intro}

In \citet{Aspden08a} (henceforth Paper~I) three-dimensional
simulations of resolved flames were performed that examined the
interactions between turbulence and a carbon-burning flame in a Type Ia
supernova at different densities.  Because of the strong dependence of
flame width and speed on density, this study can be viewed as a survey
of the behavior of the flame at variable Karlovitz number,
\begin{equation}
\Ka=\sqrt{\frac{\check{u}^3}{s_L^3}\frac{l_L}{l}}.
\end{equation}
Here $s_L$ and $l_L$ are the laminar flame speed and width,
respectively, and $\check{u}$ and $l$ are the turbulent intensity (rms
velocity) and integral length scale (defined conventionally as the
integral of the longitudinal correlation function). For $\Ka\ltaprx1$,
the flame is laminar with propagation determined by a balance 
between conductive and burning time scales. Such laminar flames have a
large Lewis number, which is to say thermal diffusion occurs much
faster than carbon diffusion.  This means that perturbing the flame
surface into the ash leads to a focusing of heat by diffusion,
enhancing the burning rate, burning away the perturbation.
Similarly, perturbing the flame surface into the fuel leads to a
defocusing of heat by diffusion, decreasing the
burning rate.  As a result, the flames are thermodiffusively stable.
At small-to-moderate Karlovitz numbers ($\Ka\ltaprx 1$), it was found
that this thermodiffusively-stable nature led to a balance between
local enhancement of the flame and local extinction, and so the
turbulent flame speed remained close to the laminar value.

However, once the Karlovitz number was sufficiently large ($\Ka
\gtaprx 10$), turbulence was sufficiently strong that the turbulent
mixing dominated thermal diffusion, and the flame became completely
stirred (it resembled a turbulent mixing zone) and its width was
greatly broadened. The turbulent flame width was also much 
greater than the integral length scale.  While the local burning rate
was greatly reduced, the overall flame speed was a factor of five or
six times the laminar flame speed due to the enhanced volume of the
burning.

\begin{figure}
\centering
\plotone{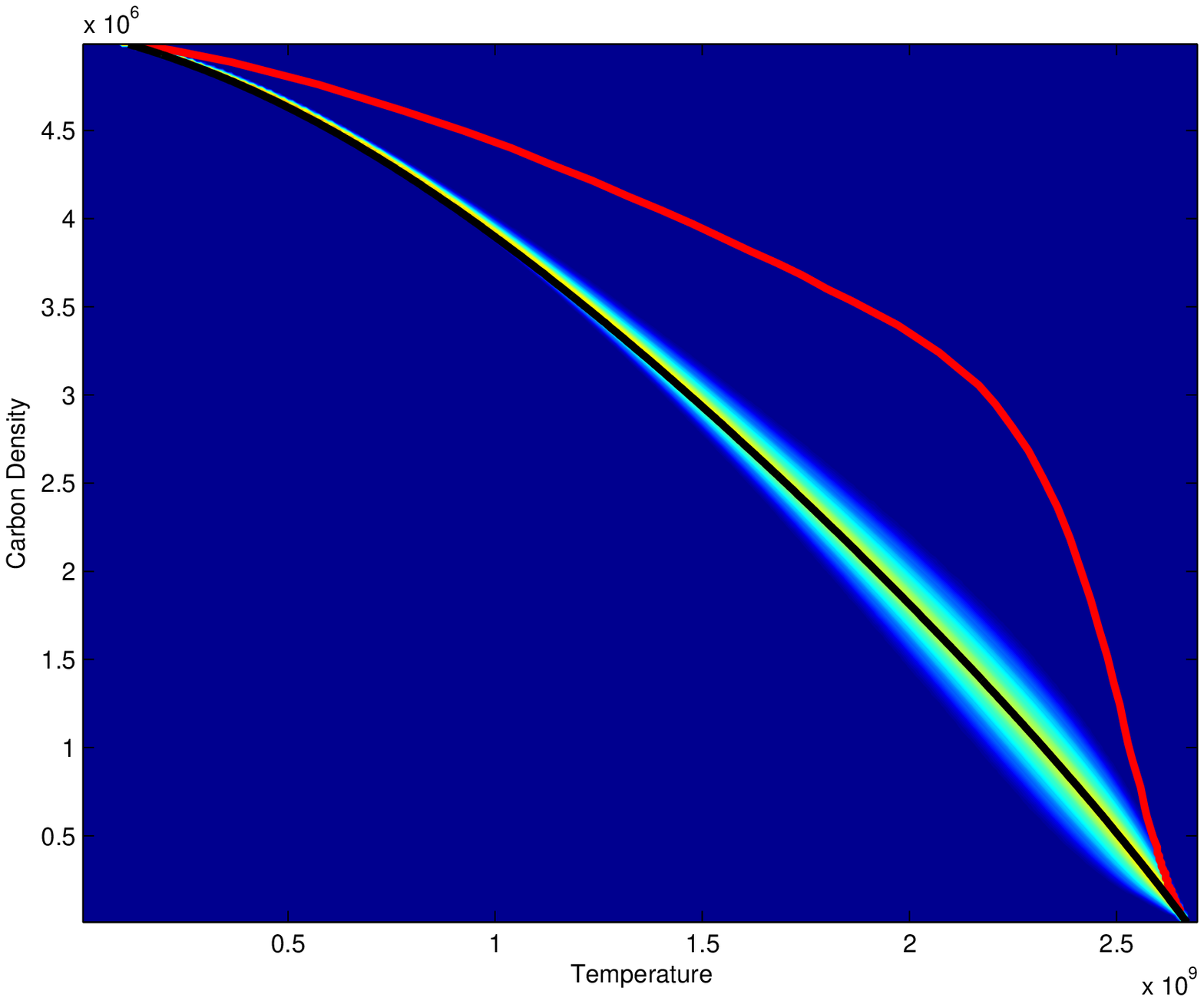}
\caption{Joint probability density function of fuel and temperature
  from case (e) in Paper~I. The solid red line shows the distribution
  from the flat laminar flame where thermal diffusion is the dominant
  mixing process.  The solid black line shows the distribution of fuel
  burning isobarically with no thermal or species diffusion.}
\label{Fig:PdfFuelTemp}
\end{figure}

A key result from Paper~I was that at high Karlovitz number
($\Ka\approx230$), turbulence dominates the mixing of fuel and heat
while thermal diffusion plays only a minor role. Figure
\ref{Fig:PdfFuelTemp} shows a joint probability density function
(\jpdf) of fuel and temperature from Paper~I. 
The solid red line shows the distribution from the flat laminar flame
where thermal diffusion is the dominant mixing process, and the solid
black line shows the distribution of fuel burning isobarically with no
thermal or species diffusion.  The agreement with the
\jpdf\ demonstrates that the turbulent flame is burning at an
effective Lewis number close to unity - the mixing is chiefly due
to turbulence.

The simulations in Paper~I were performed in small domains to ensure
that the laminar flame was fully-resolved -- the domain size was
approximately twenty-five times the laminar flame width.  The aim of this
paper is to investigate turbulent flame speeds in larger domains, with
the aim of predicting high Karlovitz flame speeds in a full star, and
specifically testing predictions made in Paper~I.  In this paper, we
will refine these predictions by giving an in-depth theoretical
description of the distributed burning regime, which are then compared
with one- and three-dimensional simulations.

\section{THEORETICAL DESCRIPTION OF THE DISTRIBUTED BURNING REGIME}

There is diversity in the literature when naming the burning
regimes for large Karlovitz number.  Here we use the term
``distributed burning regime'' (or ``distributed reaction zone'') to
refer collectively to all burning with large Karlovitz number ($\Ka
\gtaprx 10$).  We also follow \citet{Woosley09} in further subdividing
this region based upon the turbulent Damk\"ohler number, $\Da_T$, to
be defined below.  Specifically, $\Da_T<1$ will be referred to as the
``well-stirred reactor'' (see \cite{Peters86} or \cite{Woosley09}) and
$\Da_T>1$ as the ``stirred-flame regime'' (see \cite{Kerstein01} or
\cite{Woosley09}).

\subsection{The Well-Stirred Reactor}
\label{sec:wsr}

\cite{Damkohler40} identified two burning regimes, which he referred
to as ``small-scale'' and ``large-scale'' turbulence, see also
\cite{Peters99,Peters00}.  In the large-scale turbulence regime, laminar
flames are dragged around by large turbulent eddies, which is an
appropriate description of the burning in a Type Ia supernovae at high 
density \citep{Woosley09}.  The high
Karlovitz number case  from Paper~I is in the ``small-scale''
turbulence regime, where the small-scale mixing is dominated by
turbulence, rather than diffusion.  

In this regime, the turbulent eddies are strong enough to disrupt the
internal structure of the flame.  The time scale of these eddies is
faster than the time scale of the flame, so the flame is mixed before
it can burn.  Therefore, the burning takes place on an inductive time
scale, or turbulent nuclear time scale, which is much slower than the
laminar nuclear time scale.
Damk\"ohler predicted (by analogy with laminar flames) that
the turbulent flame speed $s_T$ and width $l_T$ should depend upon the
turbulent diffusion coefficient and the nuclear burning time scale,
\begin{equation}
s_T=\sqrt{\frac{\mathcal{D}_T}{\tau^T_{\mathrm{nuc}}}},
\quad\mathrm{and}\quad
l_T=\sqrt{\mathcal{D}_T\tau^T_{\mathrm{nuc}}},
\end{equation}
where $\tau^T_{\mathrm{nuc}}$ is the turbulent nuclear time scale 
($\tau^T_{\mathrm{nuc}}=l_T/s_T$), and $\mathcal{D}_T$ is the turbulent
diffusion coefficient $\mathcal{D}_T=\alpha\check{u}l$ (not to be
confused with $\Da_T$), where $\alpha$ is an order
one constant.  For convenience, take $\alpha=1$, which can be thought
of as absorbing the constant into the definition of the turbulent
flame width, which is ambiguous.

Keeping the Karlovitz number fixed (so the energy dissipation rate of
the turbulence $\varepsilon^*=\check{u}^3/l$ is constant) and assuming that
$\tau^T_{\mathrm{nuc}}$ is constant (i.e.\ the limiting value has been
achieved), then there is only one free parameter.  This parameter can
be written as a turbulent Damk\"ohler number, defined as
\begin{equation}
\Da_T\equiv\frac{\tau_T}{\tau^T_{\mathrm{nuc}}}=\frac{s_T}{\check{u}}\frac{l}{l_T}.
\end{equation}
It is straightforward to show that $\Da_T\propto l^{2/3}$, where the
constant of proportionality is
$({\varepsilon^*}^{1/3}\tau^T_{\mathrm{nuc}})^{-1}$.  Therefore, the
free parameter can be thought of in terms of the integral length
scale.  Both $s_T$ and $l_T$ are proportional to $\mathcal{D}_T^{1/2}$, so
writing $\mathcal{D}_T=\check{u}l={\varepsilon^*}^{1/3}l^{4/3}$, it follows
immediately that both $s_T$ and $l_T$ scale with $l^{2/3}$ (or
equivalently $\Da_T$).

\subsection{The Stirred-Flame Regime}

When the turbulence time scale $\tau_T$ becomes comparable to the
turbulent nuclear time scale $\tau^T_{\mathrm{nuc}}$,
i.e. $\Da_T\approx1$, the turbulence on the integral length scale can
no longer mix the flame before it burns.  Therefore, the flame can not
be broadened any further and a limiting behavior is reached.
Specifically, for $\Da_T\gtaprx1$, the flame burns like a unity Lewis
number flame (on the scale of the flame width) with local
flame speed $s_\lambda$ and width $l_\lambda$.  We refer to this kind
of burning as a $\lambda$-flame.

Defining a turbulent Karlovitz number as 
\begin{equation}
\Ka_T=\sqrt{\frac{\check{u}^3}{s_T^3}\frac{l_T}{l}},
\end{equation}
it can be shown that $\Da_T^2\Ka_T^2=\alpha\equiv1$.
Therefore, when the scaling relations break down at $\Da_T\approx1$,
it follows immediately that the turbulent flame speed is equal to the 
turbulent intensity and the turbulent flame width is equal to the
integral length scale.  

It is the limit $\Da_T\approx1$ that divides the distributed regime
into the well-stirred reactor regime and the stirred-flame regime.
In particular, note that it is the {\it turbulent} Damk\"ohler number
that is the divide, and that $\Da_T=\sigma\Da_L$, where
$\sigma=\tau_{\rm nuc}^L/\tau_{\rm nuc}^{T}$ is the ratio of the
nuclear time scales.  Therefore, a $\lambda$-flame can only exist in
the stirred-flame regime, i.e.\ $\Da_T>1$ and $\Ka\gtaprx10$.

\begin{figure}
\centering
\plotone{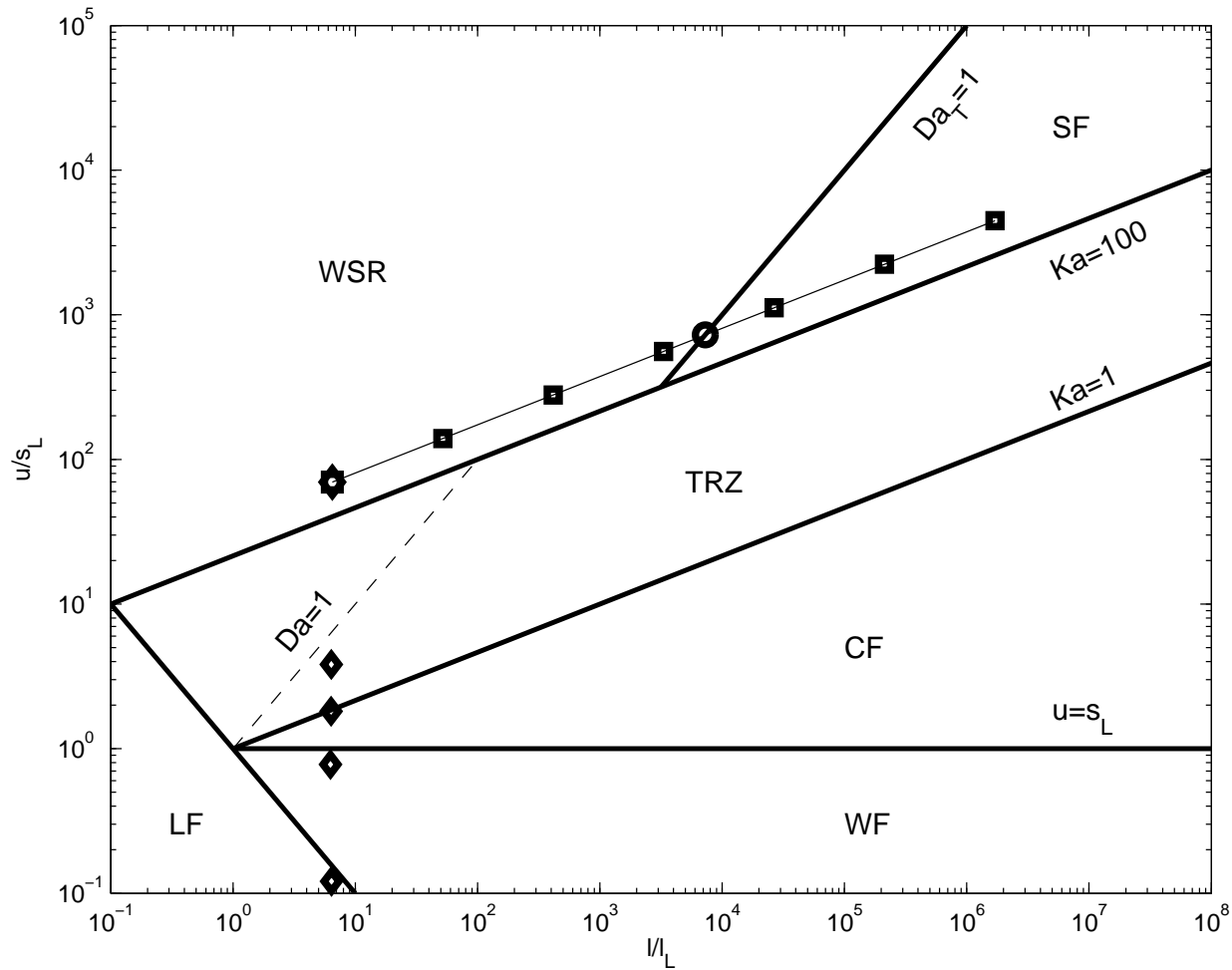}
\caption{Regime diagram based on \cite{Peters99,Peters00}.  Here we
  emphasize the separation of the distributed burning regime into the
  well-stirred reactor and stirred-flame regimes by the turbulence
  Damk\"ohler number $\Da_T=1$.  (LF - Laminar Flames, WF - Wrinkled
  Flames, CF - Corrugated Flames, TRZ - Thin Reaction Zone, WSR -
  Well-Stirred Reactor, SF - Stirred Flame).  Together the WSR and SF
  regimes make up the distributed burning regime.  The diamonds denote
  the simulations from Paper~I, and the squares denote the simulations
  from the present paper.  The circle denotes the intersect of the $\Ka=230$
  line with $\Da_T=1$, which denotes the $\lambda$-point, where the
  turbulent intensity and integral length scale are equal to the
  turbulent flame speed ($s_\lambda$) and width ($l_\lambda$),
  respectively.} 
\label{Fig:regime}
\end{figure}

Figure \ref{Fig:regime} shows a regime diagram, based on
\cite{Peters99,Peters00}, where we emphasize the divide in the
distributed burning regime.  The diamonds denote the simulations from
Paper~I, and the squares denote the simulations from the present
paper (to be defined below).  The circle denotes the intersection of
the $\Ka=230$ line with $\Da_T=1$, which denotes the $\lambda$-point,
where the turbulent intensity and integral length scale are equal to
the turbulent flame speed ($s_\lambda$) and width ($l_\lambda$),
respectively.

We emphasize that the $\lambda$-flame speed and width are local
measures, i.e.\ the flame burns at $s_\lambda$ on a scale of
$l_\lambda$.  These quantities will also vary due to turbulent
intermittency.  The $\lambda$-flame will respond to the local
turbulence, specifically $\varepsilon$ on the scale of $l_\lambda$.
Note that due to this response, the turbulent Gibson scale
$l_G^T=s_\lambda^3/(\check{u}^3/l)$ is equal to the $\lambda$-flame
width, $l_G^T=l_\lambda$, and the Karlovitz number based on the
$\lambda$-flame is always one,
i.e.\ $\Ka_\lambda^2\equiv(\check{u}^3l_\lambda)/(s_\lambda^3l)\equiv1$.

The overall turbulent flame speed will be greater than $s_\lambda$ due
to enhanced flame surface area, and will resemble Damk\"ohler's
large-scale turbulence regime, due to the presence of multiple
$\lambda$-flames across an integral length scale.  Following
\cite{Peters99}, for example, the following simple expression, based
on an enhanced flame surface area, can be used to illustrate the
scaling behavior in this regime, 
\begin{equation}
\frac{s_T}{s_\lambda}=1+\frac{\check{u}}{s_\lambda}.
\end{equation}
Specifically, in the limit of high Damk\"ohler number, the turbulent
burning speed tends to the turbulent intensity,
i.e.\ $s_T\rightarrow\check{u}$ as $\Da_T\rightarrow\infty$.

In the next section, we present calculations to investigate the
turbulent flame speed as a function of Damk\"ohler number for a high
Karlovitz number flame. Specifically, we are looking for a relation of
the form
\begin{equation}
\frac{s_T}{\check{u}} = \varphi\left(\Da_T\right),
\end{equation}
for some dimensionless function $\varphi$.
Three burning regimes are to be expected.  First, for
$\Da_T\ltaprx1$, the turbulent flame speed and width are predicted to
scale with 
\begin{equation}
\frac{s_T}{\check{u}} = \Da_T^{1/2},
\quad\mathrm{and}\quad
\frac{l_T}{l} = \Da_T^{-1/2}.
\label{Eq:DaScaling}
\end{equation}
Second, for $\Da_T\approx1$, the flame is predicted to reach a
limiting behavior with speed $s_\lambda$ and width $l_\lambda$ (on the 
scale of $l_\lambda$).  Finally, for $\Da_T\gg1$, the flame is
predicted to burn as a $\lambda$-flame, where the overall flame speed
is expected to be a few times the turbulent intensity, tending towards
it as $\Da_T$ tends to infinity.

\subsection{The Turbulent Nuclear Time Scale}

Previous work has implicitly assumed that the nuclear
time scale remains unaffected by the turbulence, again see
\cite{Peters99} or \cite{Peters00}.  This was the case in
\cite{roepkehillebrandt2004}, where the relation
\begin{equation}
\frac{s_T}{s_L}\sim\left(\frac{\mathcal{D}_T}{\mathcal{D}_L}\right)^{1/2}
\sim\left(\frac{\check{u}l}{s_Ll_L}\right)^{1/2},
\label{eq:ropke}
\end{equation}
was used to derive a turbulent flame speed for a level-set method.
However, as discussed in Paper~I, due to the different distributions of
carbon and temperature, the nuclear time scale is around an order of
magnitude longer in the turbulent case.  Allowing for different
nuclear time scales, equation (\ref{eq:ropke}) becomes  
\begin{equation}
\frac{s_T}{s_L} = 
\left(\frac{\check{u}l}{s_L l_L}\right)^{1/2} \left(\frac{\tau_{\rm  nuc}^L}{\tau_{\rm nuc}^{T}}\right)^{1/2}.
\label{eq:ropkeCorrection}
\end{equation}
In Paper~I, the turbulent nuclear time scale was estimated using
$l_T/s_T$, based on an estimate of the turbulent flame width.  It was
shown that this reduces the estimate for the turbulent flame speed in
equation \ref{eq:ropkeCorrection} by a factor of approximately 
$\sigma^{1/2}\approx0.3$.  A more refined approach for estimating
$\tau_{\rm nuc}^{T}$ is given below.

A corollary to the relation $\Da_T^2\Ka_T^2=1$, is that the turbulent
nuclear time scale can be derived from a single measurement in the
well-stirred reactor regime, specifically, the turbulent flame speed
and properties of the turbulence alone. In particular, if we consider
a reference case that is easily computed (e.g.\ case (e) from
Paper~I), where the turbulent intensity is $\check{u}_0$ and integral
length scale $l_0$, then only the turbulent flame speed $s_T^0$ needs
to be measured.  Assuming that the limiting nuclear time scale
has been achieved, then the relation $\Da_T^2\Ka_T^2=1$ means that the
turbulent flame width is $l_T^0=\check{u}_0l_0/s_T^0$, and the
turbulent nuclear time scale is
$\tau_{\mathrm{nuc}}^T=l_T^0/s_T^0=\check{u}_0l_0/{s_T^0}^2$.
For case (e) in Paper~I, this gives $l_T\approx190$cm, 
$\tau_{\rm nuc}^T\approx0.0098$s, and $\sigma\approx0.067$ (which
gives $\sigma^{1/2}\approx0.26$).

\section{SIMULATION DESCRIPTION}
\label{Sec:Numerics}

\subsection{Three-Dimensional Simulations}

As in Paper~I, we use a low Mach number hydrodynamics code, adapted to
the study of thermonuclear flames, as described in \citet{SNeCodePaper}.
The advantage of this method is that sound waves are filtered out
analytically, so the time step is set by the the bulk fluid velocity
and not the sound speed.  This is an enormous efficiency gain for low
speed flames.  The input physics used in the present simulations is
largely unchanged, with the exception of the addition of Coulomb
screening, taken from the Kepler code \citep{weaver:1978}, to the
$^{12}$C($^{12}$C,$\gamma$)$^{24}$Mg reaction rate.  This yields a
small enhancement to the flame speed, and is included for
completeness.  The conductivities are those reported in
\citet{timmes_he_flames:2000}, and the equation of state is the
Helmholtz free-energy based general stellar EOS described in
\citet{timmes_swesty:2000}.  We note that we do not utilize the
Coulomb corrections to the electron gas in the general EOS, as these
are expected to be minor at the conditions considered.

The basic discretization combines a symmetric operator-split treatment
of chemistry and transport with a density-weighted approximate
projection method.  The projection method incorporates the equation of
state by imposing a constraint on the velocity divergence.  The
resulting integration of the advective terms proceeds on the time
scale of the relatively slow advective transport.  Faster diffusion
and chemistry processes are treated time-implicitly.  This integration
scheme is embedded in a parallel adaptive mesh refinement algorithm
framework based on a hierarchical system of rectangular grid patches.
The complete integration algorithm is second-order accurate in space
and time, and discretely conserves species mass and enthalpy.
The details of the adaptive incompressible flow solver can be found in
\cite{AlmBelColHowWel98}, the reacting flow solver in \cite{DayBell00}, extension
to generalized equation of state in \cite{SNeCodePaper}, and an
application to the Rayleigh-Taylor instability in type Ia SNe in
\cite{SNrt3d}.

The non-oscillatory finite-volume scheme employed here permits the use of
implicit large eddy simulation (\iles).  This technique captures the inviscid cascade
of kinetic energy through the inertial range, while the numerical error acts
in a way that emulates the dissipative physical effects on the dynamics at the grid
scale, without the expense of resolving the entire dissipation
subrange.  
An overview of the technique can be
found in \cite{GrinsteinBook07}.  \cite{Aspden08b}
presented a detailed study of the technique using the present numerical
scheme, including a characterization that allowed for an effective viscosity
to be derived.  Thermal diffusion plays a significant
role in the flame dynamics, and so is explicitly included in the model, whereas species
diffusion is significantly smaller, and so is not explicitly included.

The turbulent velocity field was maintained using the forcing
term used in Paper~I and \cite{Aspden08b}.  Specifically, a forcing 
term was included in the momentum equations consisting of a superposition of
long wavelength Fourier modes with random amplitudes and phases. The forcing term
is scaled by density so that the forcing is somewhat reduced in the ash. This
approach provides a way to embed the flame in a zero-mean turbulent background,
mimicking the much larger inertial range that these flames would experience in
a type Ia supernova, without the need to resolve the large-scale convective
motions that drive the turbulent energy cascade.  The
  effects of resolution were examined in detail in
\cite{Aspden08b}, where it was demonstrated that the
effective Kolmogorov length scale is 
approximately $0.28\Delta x$, and the integral length scale is
approximately a tenth of the domain width.  This is
particularly relevant to the present study, because turbulence is the
dominant mixing process.  This means that the \iles\ approach can be
used to capture the effects of turbulent mixing, which occurs on
length scales much larger than the actual Kolmogorov length scale in
the star.

\begin{figure}
\centering
\epsscale{0.8}
\plotone{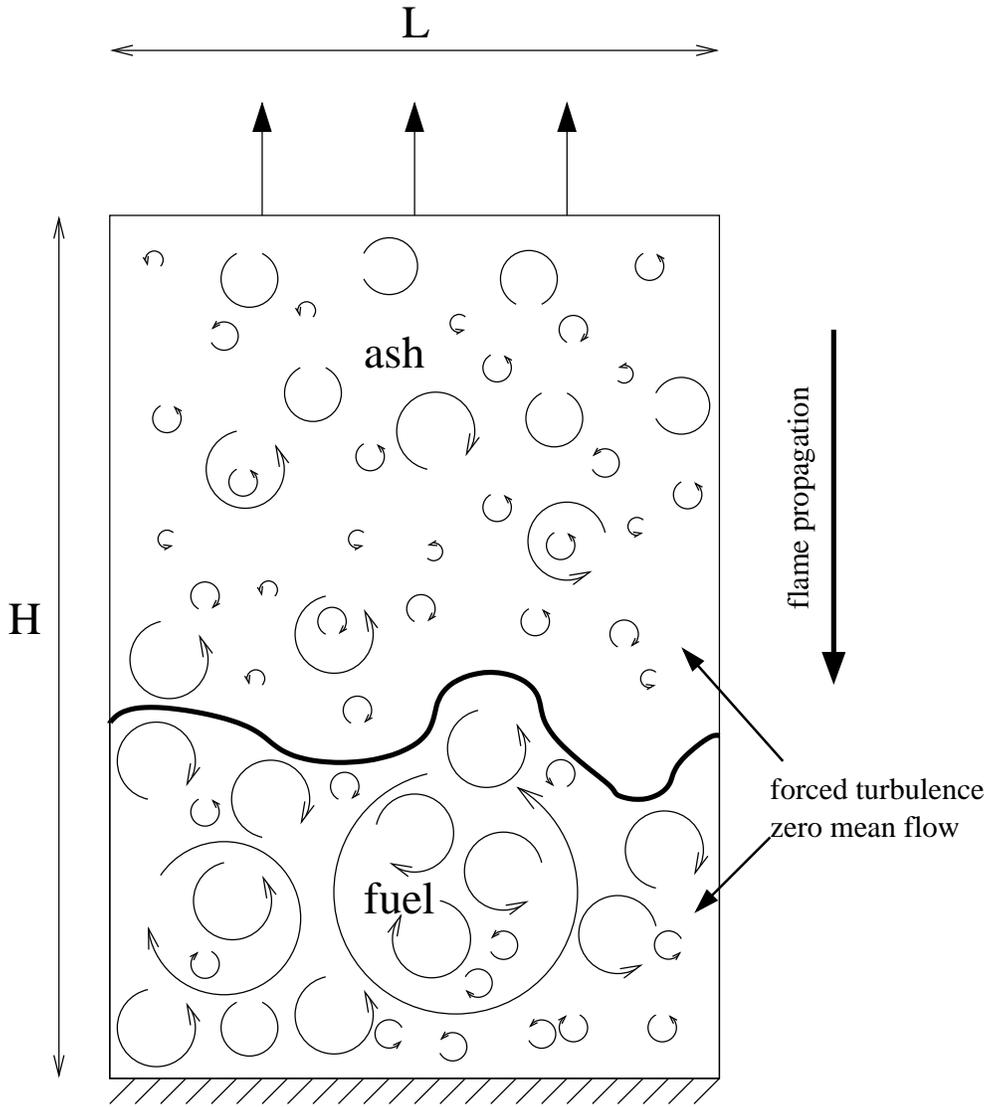}
\epsscale{1.0}
\caption{Diagram of the simulation setup (shown in
two-dimensions for clarity).
The domain is initialized with a turbulent flow and a flame is introduced into the
domain, oriented to that the flame propagates toward the lower boundary.
The turbulence is maintained by adding a forcing term to the momentum equations.
The top and bottom boundaries are outflow and solid wall boundaries, respectively.
The side boundaries are
periodic.}
\label{Fig:Setup}
\end{figure}

Figure~\ref{Fig:Setup} shows the simulation setup.  The simulations were
initialized with carbon fuel in the lower part of the domain and magnesium ash
in the upper, resulting in a downward propagating flame.  A high-aspect ratio
domain was used to allow the flame sufficient space to propagate.  Periodic
boundary conditions were prescribed laterally, along with a free-slip base,
and outflow at the upper boundary.

\subsection{One-Dimensional Simulations Using the Linear Eddy Model}

Simulations were also run using the Linear Eddy Model (LEM) of
\cite{Kerstein91}.  This approach simulates the evolution of scalar
properties in a one-dimensional domain, which can be interpreted as a
line of sight through a three-dimensional turbulent flow.  Diffusive
transport and chemical reactions are coupled in a model that
represents the effects of real three-dimensional eddies through a
so-called ``triplet map''. The advantage of LEM is that much
finer resolution and larger length scales can be explored
inexpensively. This is particularly important for the present problem,
where the range of length scales is very large and the degree of
turbulence very high. LEM has been successfully applied to a large
range of phenomena, especially turbulent terrestrial combustion.  Its
disadvantage is that when applied to a novel environment like a
supernova, there are two overall normalization factors that must be
adjusted, the effective turbulent dissipation rate and the integral
scale. By using LEM for the present problem, we hope to satisfy two
goals - first to show that very similar answers are obtained using
quite different techniques and second, to calibrate the uncertain
constants in LEM for the supernova problem.

LEM has previously been compared with the same three-dimensional code
in \citet{Woosley09}. When the nuclear physics and fuel temperature
were adjusted to be the same, it was found that best agreement
occurred for an LEM constant $C$ = 11 and an integral scale in LEM
that was three times that in the numerical simulation. Those same
values are used here except that $C$ has been assumed to be 10.  As
expected, this prescription gives excellent agreement in the
well-stirred reactor regime where it was calibrated, but, as we shall
see underestimates the flame speed by as much as a factor of two in
the stirred-flame regime. Thus a normalization that depends on $\Da_T$ is
appropriate, and $C$ = 3 to 5 for $\Da_T > 1$. 

\section{RESULTS}
\label{Sec:Results}

\subsection{Three-Dimensional Results}

The high Karlovitz case from Paper~I was used as the starting point
for the present study, and is referred to here as case (a).  Because
turbulent diffusion dominates the mixing of fuel and temperature, the
resolution requirements are significantly relaxed -- thermal diffusion
has to be resolved to capture the laminar flame correctly, but the
turbulent diffusion coefficient is much larger, and so fewer cells are
required to resolve it.  In fact, keeping $\varepsilon^*$ constant means
that the diffusion coefficient scales with $\mathcal{D}_T\sim l^{4/3}$, and a
mixing length can be defined analogous to the Kolmogorov length scale
$\eta_{\mathcal{D}}=(\mathcal{D}_T^3/\varepsilon^*)^{1/4}$, which scales as $\eta_{\mathcal{D}}\sim l$. 
Thus, the smallest scales that need to be resolved become larger as
the domain size grows.

The approach taken to achieve larger Damk\"ohler numbers (i.e.\ larger
length scales) was to start with case (a), which has a domain width of
$1.5\times10^{2}$cm, and run the same case at an eighth of the
resolution, i.e.\ with a low-resolution cell width of $\Delta x_{LR}=\Delta x_{HR}/8$.
The turbulent flame speed was used as the primary diagnostic.
Agreement in the turbulent flame speed provides confidence in
capturing the effects of turbulent mixing (relying on the \iles\ approach) with the new cell
width; it is the turbulent flame speed that is of primary interest.  A higher
Damk\"ohler number was then achieved by running in a domain eight times
larger with the new cell width, i.e.\ with a domain size
of $1.2\times10^{3}$cm.  The turbulent intensity was adjusted
accordingly to keep $\varepsilon^*=\check{u}^3/l$ constant (i.e.\ an
eight-fold increase in length scale corresponds to a two-fold increase
in velocity).  The whole process was then repeated several times to
reach a domain size of $6.14\times10^{5}$cm, and
span a range of Damk\"ohler numbers from the base case of
$\Da_T\approx0.0064$ to $\Da_T\approx1.68$, corresponding to cases (a)
through (e).

For larger length scales, the above argument no longer applies because
the diffusion coefficient responsible for mixing fuel and temperature
is approximately $s_\lambda l_\lambda$ and does not scale with
$\check{u}l$ for $\Da_T\gtaprx1$; even case (e) is questionably
resolved.  However, we have included simulations (f) and (g) at
Damk\"ohler numbers 6.6 and 26, respectively, as an indication of what
we can expect for $\Da_T\gtaprx1$.  The simulation properties are
summarized in table \ref{Tab:SimProperties}.

\begin{figure}
\centering
\plotone{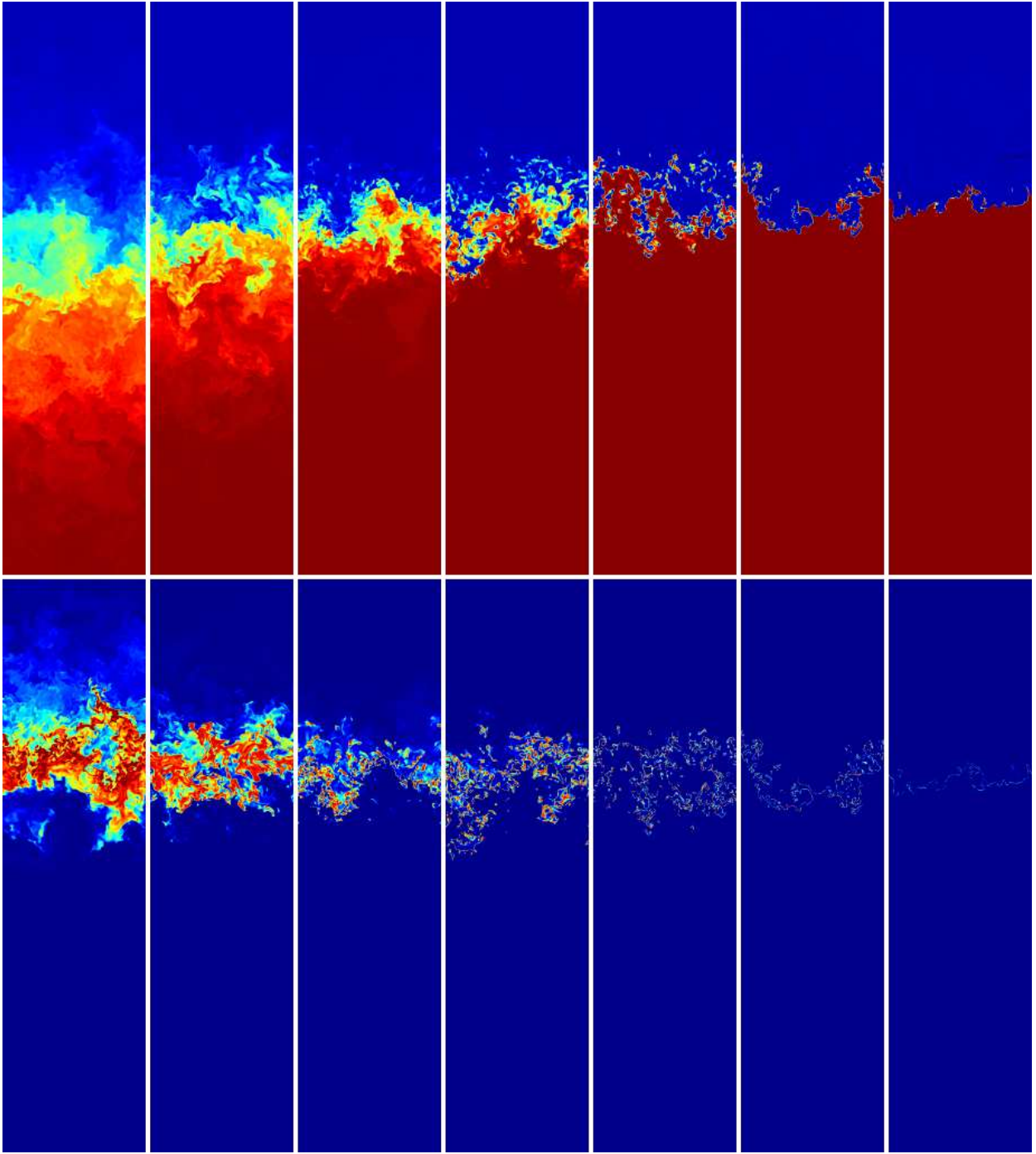}
\caption{Two-dimensional slices showing density (top) and fuel
  consumption rate (bottom) for all cases (a)-(g) (left-to-right).
  All of the figures have been normalized by the same values.
  It is important to note the domain increases by a factor of 8 each
  time.  In particular, the domain size in the final case is over two
  hundred and fifty thousand times larger than the first case; despite
  the burning rate looking reduced, it is actually greatly increased.}
\label{Fig:Panels}
\end{figure}

Figure \ref{Fig:Panels} shows slices of density (top) and fuel
consumption rate (bottom) for the seven cases (a)-(g) left-to-right.
Note that the domain size increases by a factor of 8 each time.  All
of the figures have been normalized by the same values.  Case~(a)
presents an extremely broad mixing region (much broader than the 
integral length scale), and the burning can be seen to occur at the
high temperature (low density) end of the mixing zone.  As the
Damk\"ohler number is increased, the relative width of the flame brush
decreases as expected.  Although the width of the flame brush appears
to decrease, it actually increases, just more slowly than the domain
size.  For $\Da_T\gtaprx1$, there appears to be a sharp interface
between the fuel and products.  This is because of the underresolved
nature of the flames - the actual flame will be a broad mixing zone,
but is not fully-captured here.

\begin{figure}
\centering
\plotone{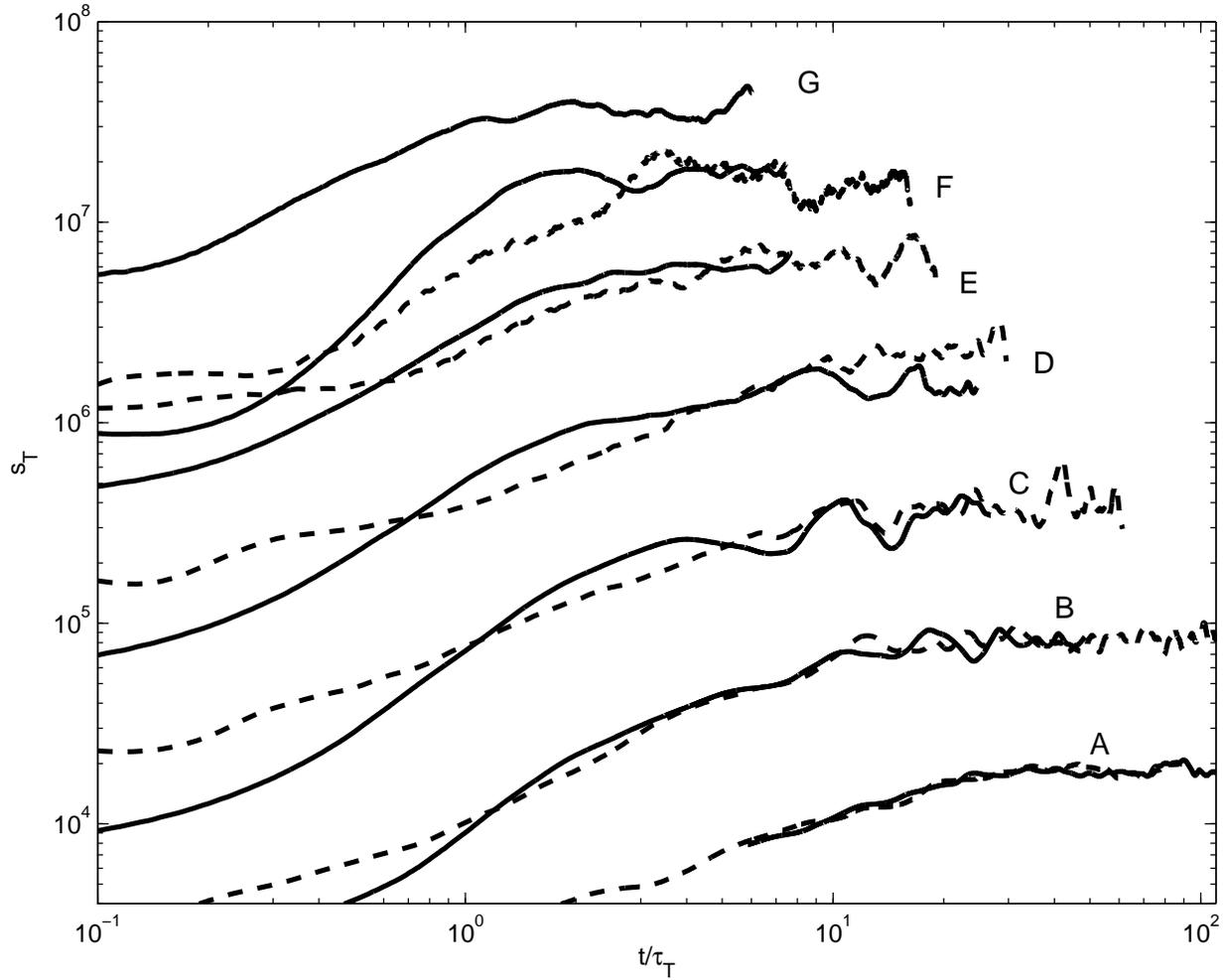}
\caption{Turbulent flame speeds $s_T$ as a function of time (normalized by
  turbulent time scale $\tau_T$.  The solid lines denote the
  simulations at the full resolution ($256^2$ cells in cross-section),
  and the dashed lines denote the simulations at the low resolution
  ($32^2$ cells in cross-section).}
\label{Fig:FlameSpeeds}
\end{figure}

Figure \ref{Fig:FlameSpeeds} shows the turbulent flame speed evolution
for the seven cases.  The flame speed is evaluated by finding the rate
of change of the total fuel mass in the domain divided by the product
of the cross-sectional area of the domain and the fuel density,
\begin{equation}
s_T=\frac{1}{A(\rho X_C)_0}\frac{\D}{\D t}\int_V{\rho X_C \,\D V}.
\end{equation}
The time scale has been normalized by the eddy turnover time
$\tau_T=l/\check{u}$.  The solid lines denote the simulations at the
full resolution ($256^2$ cells in cross-section), and the dashed lines
denote the simulations at the low resolution ($32^2$ cells in
cross-section).  In most cases the low resolution simulations are in good
agreement with the high resolution simulations (the only real outlier
is case (d) where the low resolution simulation over-predicts the flame
speed by 48\%).  Table \ref{Tab:Speeds} shows the mean flame speeds at
both resolutions in each case.  

\begin{figure}
\centering
\plotone{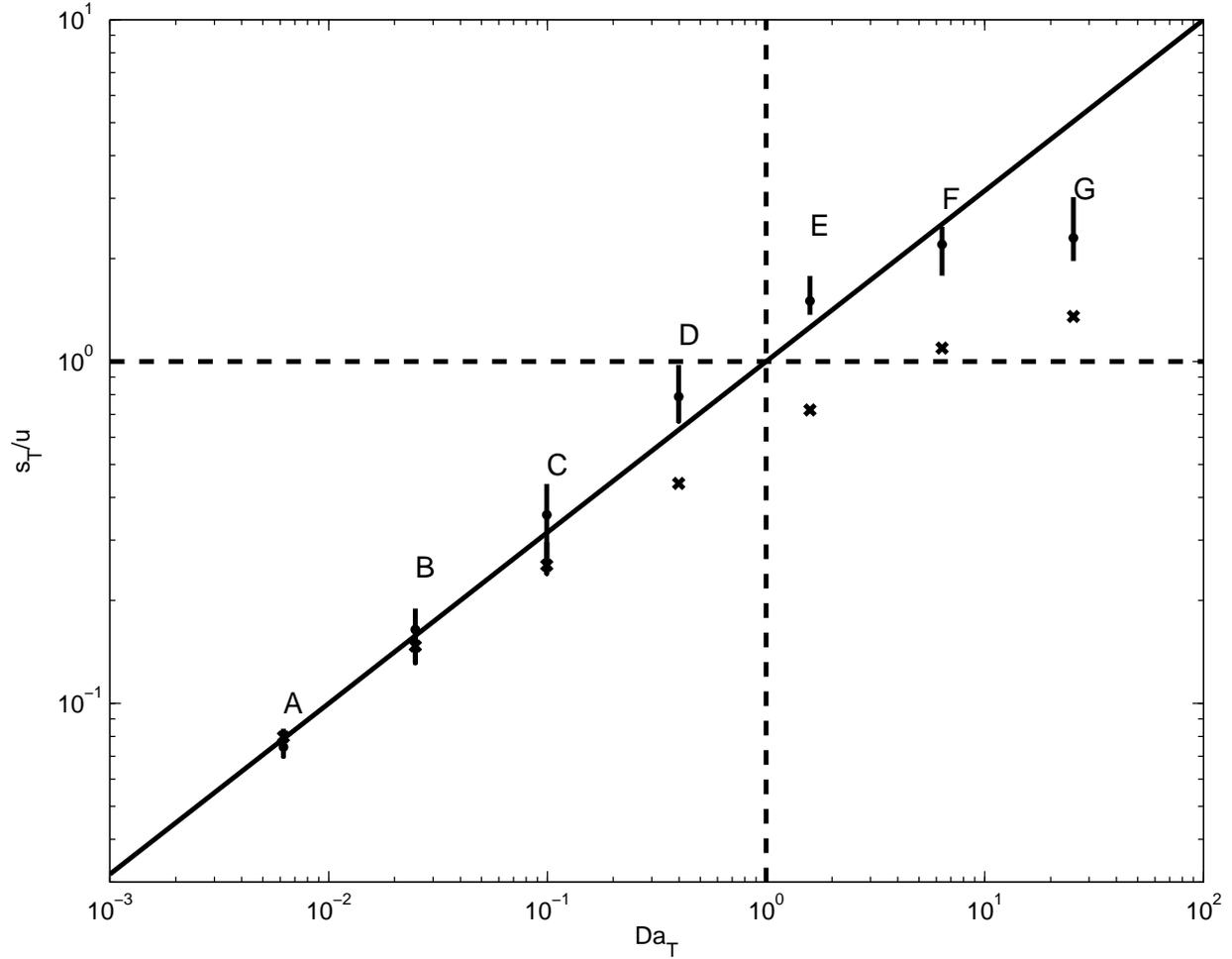}
\caption{Turbulent flame speeds normalized by turbulent intensity
  $s_T/\check{u}$ as a function of turbulent Damk\"ohler number
  $\Da_T$.  The solid black line denotes the expected scaling
  behavior for $\Da_T\ltaprx1$, bounded by the dashed lines.
  The flame speeds appear to be in good agreement with the predicted
  scaling, and despite the lack of resolution, appear to roll over when
  $\Da_T\gtaprx1$.  The crosses denote the mean speeds from the LEM
  calculations.}
\label{Fig:DaScaling}
\end{figure}

Figure \ref{Fig:DaScaling} shows the turbulent burning speeds
normalized by the turbulent intensity as a function of Damk\"ohler
number.  Case (a) is the low Damk\"ohler number on the left,
increasing to the right to case (g).  The marker denotes the mean in
each case, and the vertical line denotes the range of values obtained
during the averaging period (note that it is not the standard deviation).
The solid line is the expected scaling relation from equation
\ref{Eq:DaScaling}.  The dashed lines denote $\Da_T=1$ and
$s_T=\check{u}$, where the scaling relation is predicted to break
down.  Cases (a)-(e) are in very good agreement with the predicted
scaling.  We note that for cases (d) and (e), the flame speed 
measured is enhanced by the area of the flame that is burning, which
can be increased by turbulence here because the flame brush is now
smaller than the domain width; this explains the higher speeds
obtained.  The flames in these cases are too convoluted to extract a
flame area to normalize by.  Despite being significantly underresolved,
cases (f) and (g) appear to show the expected break-down of the
Damk\"ohler scaling; for $\Da_T\gtaprx1$, the turbulence cannot
enhance the flame speed according to equation (\ref{Eq:DaScaling}), and
the normalized flame speed ceases to grow.

\begin{figure}
\centering
\plottwo{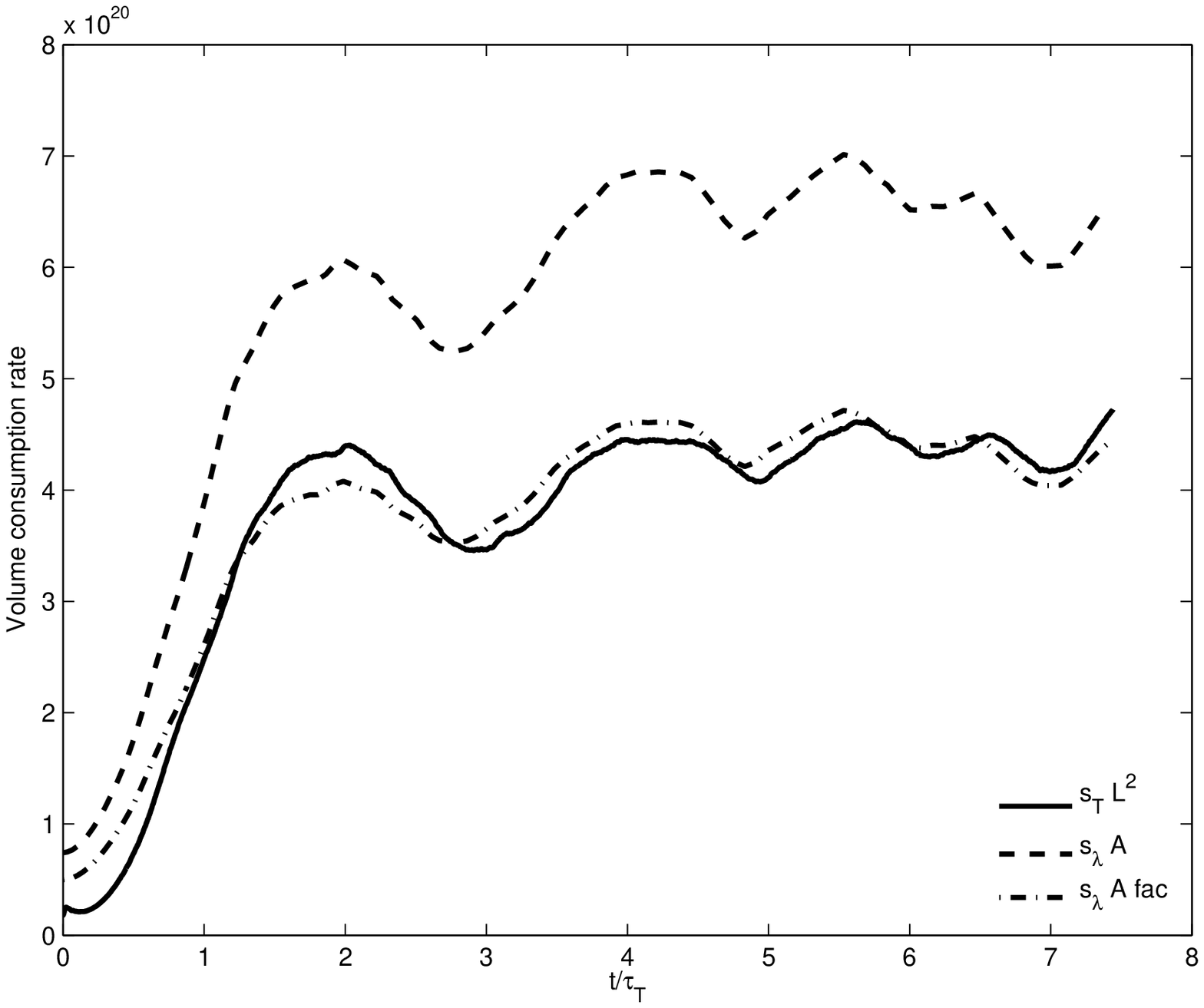}{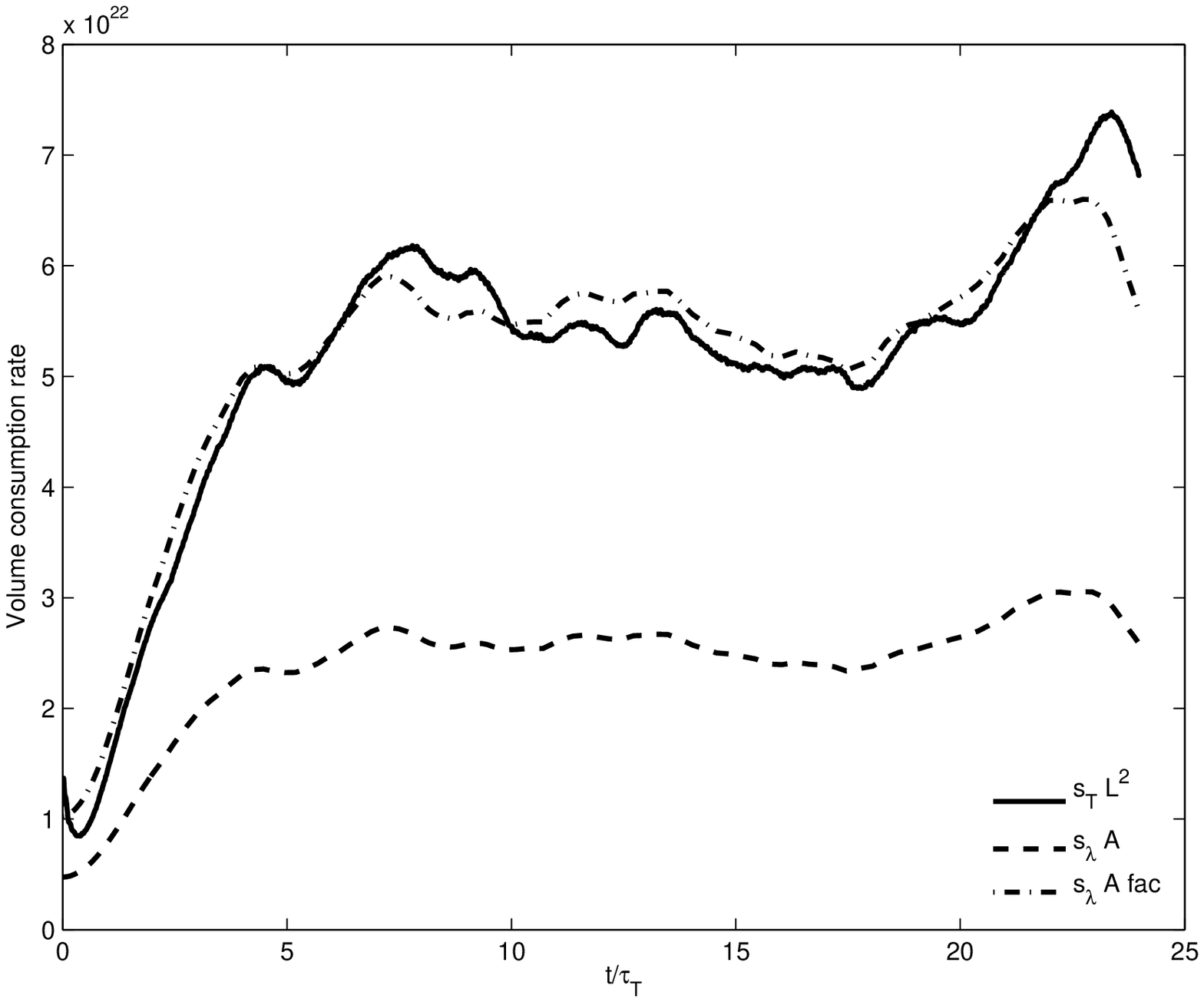}
\caption{Volumetric rate of fuel consumption.  The solid line is the 
  measured rate $As_T$, where $A$ is the cross-sectional area, the
  dotted line is the product of the limiting flame speed $s_\lambda$ with the
  measured flame surface area, and the dash-dotted line is the latter
  adjusted by a factor of $0.67$ for case (f) and $2.16$ for case
  (g).  There is a definite correlation between the flame surface area
  and the fuel consumption, but the estimate $s_\lambda$ appears to be
  an over-prediction for case (f) and an under-prediction for case (g). 
  We speculate that this is due to the underresolved nature of these
  cases and that $s_\lambda$ is still a reasonable estimate for a
  turbulent flame model.}
\label{Fig:Area}
\end{figure}

Figure \ref{Fig:Area} compares the volumetric rate of burning for
cases (f) and (g).  Because these two cases are much less convoluted
than cases (a)-(e) it was possible to extract an isosurface based on a
temperature of $10^9$K.  The solid line denotes the measured turbulent
flame speed multiplied by the cross-sectional area of the domain, the
dashed line denotes the measured flame surface area times the predicted
burning speed $s_\lambda$.   It appears that the flame speed is
over-predicted in case (f) and under-predicted in case (g); the dash-dotted
line denotes this speed multiplied by a factor of $0.67$ for case (f)
and $2.16$ for case (g).  We speculate that the disagreement is due to
the underresolved nature of these cases, but maintain that $s_\lambda$
is still a reasonable estimate for a turbulent flame model.

\subsection{One-Dimensional Results}

LEM was used to simulate the same conditions as in Table
\ref{Tab:SimProperties} except that in each case the integral scale
was multiplied by three and finer resolution was employed (Table
\ref{Tab:LEM}). Roughly 40,000 time points were sampled for each
case. The average flame speeds are plotted in 
Figure \ref{Fig:DaScaling}. For $\Da_T \ltaprx 1$, the agreement with
the three-dimensional simulations is excellent. Both show the same
scaling relation as well as agreeing on the actual value of the speed.
However, at large values of $\Da_T$, the LEM results are almost a factor
of two smaller. There could be several reasons for this. First, there
are fundamental differences in the two approaches. LEM does not
capture all of the multi-dimensional effects and has not been calibrated
for this regime. Second, the resolution of the three-dimensional study
is low and not all burning structures are well resolved. Studies with
LEM suggest a mild dependence of $s_T$ on the resolution with slower
speeds at higher resolution.

\subsection{Variability of the Turbulent Flame Speed}

\begin{figure}
\centering
\plotone{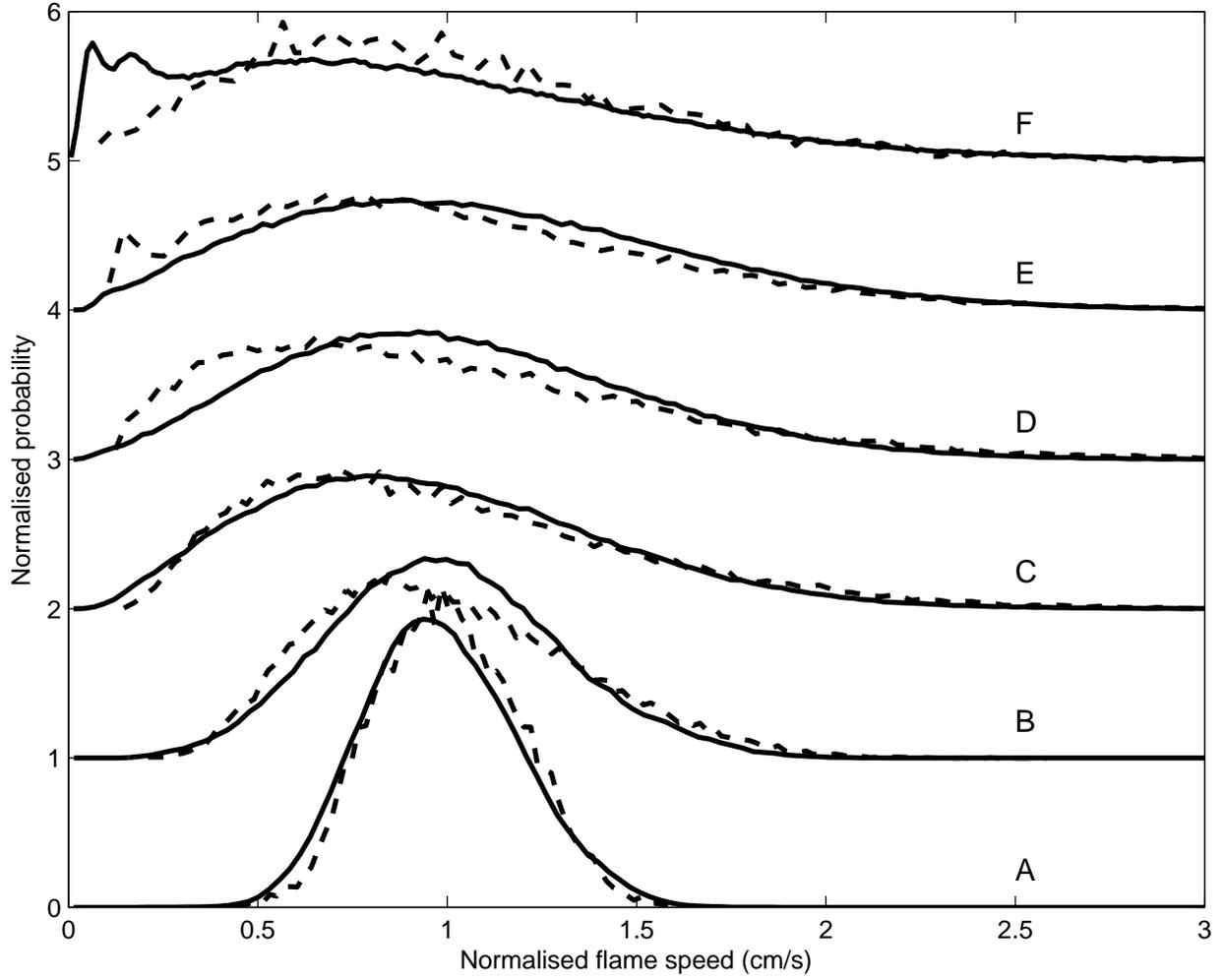}
\caption{Probability density functions of normalized local burning
  rate for both the 3D (solid) and LEM (dashed) calculations.  Each
  case has been shifted along the $y$-axis for clarity.}
\label{Fig:blockPdf}
\end{figure}

Figure \ref{Fig:blockPdf} compares the probability density functions
(\pdf s) of the normalized local burning rate in the three-dimensional
and LEM calculations.  To compare the three-dimensional results as
closely as possible with LEM, the \pdf s were evaluated by integrating
the fuel consumption along line-outs, i.e.\ over individual columns of
data as a function of height.
Each \pdf\ was then normalized by the mean burning speed.  The \pdf s have
been shifted along the $y$-axis for clarity.  In case (a), the \pdf\ has a
narrow Gaussian distribution centered around the mean flame speed.  As
the Damk\"ohler number increases, the \pdf\ becomes broader, i.e.\ a
greater range of burning rates is observed locally.

Both studies show a strong dependence of the spread of the \pdf\ on
Damk\"ohler number. In the well-stirred reactor ($\Da_T < 1$), the
integral scale is less than the flame width. Many eddies turn over on
the largest scale as burning moves through a region. The relation
between temperature, carbon mass fraction, and location is smooth and
the burning time well defined. There is only one flame
structure. Consequently the speed does not vary greatly from the
average.

In the stirred flame, however, there are multiple regions of
burning. Sometimes the burning is very fast, especially when a large
single eddy envelops a new region of fuel. At other times it, almost
goes out. The \pdf\ thus shows a large spread in speed.  Occasionally,
the overall burning proceeds at a rate that is 2.5 to 3 times the
average. Since the average itself if roughly twice $\check u$, this
implies an overall burning rate up to six times $\check u$.

\section{CONCLUSIONS}
\label{sec:conclusions}

New one- and three-dimensional simulations have been presented to
clarify and quantify the nature of carbon burning in a Type Ia
supernova in the distributed burning regime. The characteristics of
distributed burning depend upon the Damk\"ohler number, and a range of
$\Da_T$ from 0.006 to 26 has been explored.  For $\Da \ltaprx 1$,
the expected scaling relations were demonstrated,
\begin{equation}
\frac{s_T}{\check{u}} = \Da_T^{1/2},
\quad\mathrm{and}\quad
\frac{l_T}{l} = \Da_T^{-1/2}.
\label{eq:concRelations}
\end{equation}

For $\Da_T \gtaprx 1$, these relations break down and the turbulent
flame reaches a maximum local flame speed $s_\lambda$ and width
$l_\lambda$, measured on the scale of $l_\lambda$.  This
$\lambda$-flame interacts with turbulent eddies with length scales
between $l_\lambda$ and the integral length scale  $l$, which enhances
the flame surface area.  The average turbulent flame properties are
constrained by the large scales  \citep[see also][]{Woosley09}.  This
leads to an average overall turbulent flame speed, which to
factor-of-two accuracy is given by the turbulent intensity, $\check
u$. For the range of $\Da \gtaprx 1$ studied here, a good
approximation is $s_T = 2 \check u$. 

\subsection{Consequences for Large-Scale Simulations}

Large-scale simulations (i.e.\ of a full-star) will require a
turbulent flame model.  One of the consequences of the results
presented here is that the $\lambda$-flame is ideally suited to the
level-set approach, as it burns locally with speed $s_\lambda$.  This
means that a turbulent flame speed can be evaluated in the following
manner.  Assume that the turbulence is sufficiently large and intense,
i.e.\ in the stirred-flame regime ($\Da_T>1$ and $\Ka\gtaprx10$).
Then the turbulence can be characterized by the energy dissipation
rate $\varepsilon^*=\check{u}^3/l$, and the burning depends on the
turbulent nuclear burning time scale $\tau_{\mathrm{nuc}}^T$.  The
nuclear burning time scale is equal to $l_\lambda/s_\lambda$ and the
turbulent intensity at the length scale of the $\lambda$-flame is
$s_\lambda$ and so $\varepsilon^*=s_\lambda^3/l_\lambda$.  Solving for
$s_\lambda$ and $l_\lambda$ gives
\begin{equation}
s_\lambda = \sqrt{\varepsilon^*\tau_{\mathrm{nuc}}^T},
\quad\mathrm{and}\quad
l_\lambda = \sqrt{\varepsilon^*{\tau_{\mathrm{nuc}}^T}^3}.
\label{eq:lambda}
\end{equation}
Note these expressions do not require a fixed Karlovitz number
(it's incorporated by $\varepsilon^*$), and the the length scale is the
same as predicted in Paper I.  To implement a level-set approach for
a $\lambda$-flame, $\tau_{\mathrm{nuc}}^T$ should be evaluated as a
function of fuel density and temperature (using the approach outlined
in section \ref{sec:wsr} or \cite{Woosley09}) and coupled with the
(local) energy dissipation rate to evaluate the $\lambda$-flame speed
according to equation (\ref{eq:lambda}).  If the resolution requirements of
Paper~I were applied to the $\lambda$-flame recovered in
the present paper (i.e.\ $l_\lambda\approx4\Delta x$ with $256^2$
cells in cross-section), an integral length scale could be achieved
that is of order thousands of times larger than the original study.

\subsection{Consequences for Detonation}

The instantaneous flame speed for $\Da_T \gtaprx 1$ is irregular,
with frequent excursions up to 2.5 times the average, which is
approximately twice the turbulent intensity for the simulations
presented here. These variations could be crucial if a transition to
detonation is to occur. Detonation will not happen in the laminar
regime, i.e.\ at high density, because each flamelet burns at a steady
rate and has a thickness that is far below the critical mass for
detonation. It is also unlikely to happen in the well-stirred reactor
regime, because the burning is steady and slower than the turbulence
on the integral scale, which in turn is subsonic.  The burning time
scale is very long in the well-stirred regime, making it very
difficult to burn, for example, a 10 km region on a sound crossing
time.

Thus if detonation is to occur spontaneously, it is likely to happen
in the stirred-flame regime.  Even there though, $\Da_T$ should not be
too large, or the mixed regions will  not burn supersonically (recall
$s_T\rightarrow\check{u}$ for large $\Da_T$). The best conditions
therefore occur where there are multiple $\lambda$-flames across an
integral length scale, i.e.\ for small $\Da_T \gtaprx 1$
\citep{Woosley07,Woosley09}. The calibration of 
LEM determined by comparison with the three-dimensional studies
suggest a normalization constant of $C = 5$ is most appropriate for
$\Da > 1$ and this is the value used by \citet{Woosley09}. This
strengthens their conclusion that a detonation is possible if
$\check{u}$ exceeds about one-fifth sonic.

\acknowledgements

Support for A.~J.~A.\ was provided by a Seaborg Fellowship
at Lawrence Berkeley National Laboratory under Contract No. DE-AC02-05CH11231.
The work of J.~B.~B. was supported by the Applied Mathematics Research
Program of the U.S. Department of Energy under Contract No. DE-AC02-05CH11231.
At UCSC this research has been supported by the NASA Theory
Program NNG05GG08G and the DOE SciDAC Program (DE-FC02-06ER41438).
The computations presented here were performed on the ATLAS Linux
Cluster at LLNL as part of a Grand Challenge Project. 

\bibliographystyle{apj}
\bibliography{paper}

\clearpage

\clearpage

\clearpage

\clearpage

\clearpage

\clearpage

\clearpage

\clearpage

\clearpage

\begin{landscape}
\begin{table}
\begin{tabular}{|l||c|c|c|c|c|c|c|}
\hline
Case & (a) & (b) & (c) & (d) & (e) & (f) & (g) \\
\hline
Domain width, $L$ (cm) & $1.5\times10^{2}$ & $1.2\times10^{3}$ & $9.6\times10^{3}$ & $7.68\times10^{4}$ & $6.14\times10^{5}$ & $4.92\times10^{6}$ & $3.93\times10^{7}$ \\
Domain height, $H$ (cm) & $1.2\times10^{3}$ & $4.8\times10^{3}$ & $3.84\times10^{4}$ & $3.07\times10^{5}$ & $2.46\times10^{6}$ & $1.97\times10^{7}$ & $1.57\times10^{8}$ \\
Integral length scale, $l$ (cm) & $1.5\times10^{1}$ & $1.2\times10^{2}$ & $9.6\times10^{2}$ & $7.68\times10^{3}$ & $6.14\times10^{4}$ & $4.92\times10^{5}$ & $3.93\times10^{6}$ \\
Turbulent intensity, $\check{u}$ (cm/s) & $2.47\times10^{5}$ & $4.93\times10^{5}$ & $9.86\times10^{5}$ & $1.97\times10^{6}$ & $3.95\times10^{6}$ & $7.89\times10^{6}$ & $1.58\times10^{7}$ \\
Laminar Damk\"ohler, $\Da_L$ & $9.32\times10^{-2}$ & $3.73\times10^{-1}$ & $1.49\times10^{0}$ & $5.97\times10^{0}$ & $2.39\times10^{1}$ & $9.54\times10^{1}$ & $3.82\times10^{2}$ \\
Turbulent Damk\"ohler, $\Da_T$ & $6.23\times10^{-3}$ & $2.49\times10^{-2}$ & $9.97\times10^{-1}$ & $3.99\times10^{-1}$ & $1.60\times10^{0}$ & $6.38\times10^{0}$ & $2.55\times10^{1}$ \\
High resolution ($N=256$) $\Delta x_{HR}$ (cm) & $5.86\times10^{-1}$ & $4.69\times10^{0}$ & $3.75\times10^{1}$ & $3.00\times10^{2}$ & $2.40\times10^{3}$ & $1.92\times10^{4}$ & $1.54\times10^{5}$ \\
Low resolution ($N=32$) $\Delta x_{LR}$ (cm) & $4.69\times10^{0}$ & $3.75\times10^{1}$ & $3.00\times10^{2}$ & $2.40\times10^{3}$ & $1.92\times10^{4}$ & $1.54\times10^{5}$ & $1.23\times10^{6}$ \\
\hline
\end{tabular}
\caption{Simulation properties.}
\label{Tab:SimProperties}
\end{table}
\end{landscape}

\clearpage
\begin{landscape}
\begin{table}
\begin{tabular}{|l||c|c|c|c|c|c|c|}
\hline
Case & (a) & (b) & (c) & (d) & (e) & (f) & (g) \\
\hline
High resolution (cm/s) & $1.84\times10^{4}$ & $8.10\times10^{4}$ & $3.51\times10^{5}$ & $1.56\times10^{6}$ & $5.94\times10^{6}$ & $1.66\times10^{7}$ & $3.49\times10^{7}$ \\
Low resolution (cm/s) & $1.86\times10^{4}$ & $8.57\times10^{4}$ & $3.91\times10^{5}$ & $2.31\times10^{6}$ & $6.5\times10^{6}$ & $1.62\times10^{7}$ & $3.16\times10^{7}$ \\
Percentage error & 1.1 & 5.8 & 11.4 & 48.1 & 9.4 & -2.4 & -9.5 \\
\hline
\end{tabular}
\caption{Measured turbulent flame speeds.}
\label{Tab:Speeds}
\end{table}
\end{landscape}

\clearpage

\begin{landscape}
\begin{table}
\begin{tabular}{|l||c|c|c|c|c|c|c|}
\hline
Case & (a) & (b) & (c) & (d) & (e) & (f) & (g) \\
\hline
Turbulent intensity, $\check{u}$ (cm/s) & $2.47\times10^{5}$ & $4.93\times10^{5}$ & $9.86\times10^{5}$ & $1.97\times10^{6}$ & $3.95\times10^{6}$ & $7.89\times10^{6}$ & $1.58\times10^{7}$ \\
Integral length scale, $l$ (cm) & $4.50\times10^{1}$ & $3.60\times10^{2}$ & $2.88\times10^{3}$ & $2.30\times10^{4}$ & $1.84\times10^{5}$ & $1.48\times10^{6}$ & $1.18\times10^{7}$ \\
Resolution, $\Delta x$ (cm) & $4.89\times10^{-1}$ & $2.93\times10^{0}$ & $1.47\times10^{1}$ & $9.77\times10^{1}$ & $7.32\times10^{2}$ & $2.44\times10^{3}$ & $2.44\times10^{4}$ \\
Average speed (cm/s) & $1.96\times10^{4}$ & $7.26\times10^{4}$ & $2.50\times10^{5}$ & $8.70\times10^{5}$ & $2.85\times10^{6}$ & $8.36\times10^{6}$ & $2.14\times10^{7}$ \\
\hline
\end{tabular}
\caption{Characteristics of LEM studies.}
\label{Tab:LEM}
\end{table}
\end{landscape}

\end{document}